\documentclass[aip,jap,reprint]{revtex4-1}

\usepackage{graphicx}
\usepackage{epstopdf}

\begin{document}


\title{Characterization of a multimode coplanar waveguide parametric amplifier} 



\author{M. Simoen}
\email[]{simoen@chalmers.se}
\affiliation{Department of Microtechnology and Nanoscience (MC2), Chalmers University of Technology, SE-412 96 G\"{o}teborg, Sweden}
\author{C. W. S. Chang}
\affiliation{Institute for Quantum Computing and Department of Electrical and Computer Engineering, University of Waterloo, Waterloo ON N2L 3G1, Canada}
\author{P. Krantz}
\affiliation{Department of Microtechnology and Nanoscience (MC2), Chalmers University of Technology, SE-412 96 G\"{o}teborg, Sweden}
\author{Jonas Bylander}
\affiliation{Department of Microtechnology and Nanoscience (MC2), Chalmers University of Technology, SE-412 96 G\"{o}teborg, Sweden}
\author{W. Wustmann}
\affiliation{Laboratory for Physical Sciences, College Park, MD 20740, USA}
\author{V. Shumeiko}
\affiliation{Department of Microtechnology and Nanoscience (MC2), Chalmers University of Technology, SE-412 96 G\"{o}teborg, Sweden}
\author{P. Delsing}
\affiliation{Department of Microtechnology and Nanoscience (MC2), Chalmers University of Technology, SE-412 96 G\"{o}teborg, Sweden}
\author{C. M. Wilson}
\affiliation{Institute for Quantum Computing (IQC), University of Waterloo, Waterloo ON N2L 3G1, Canada}


\date{\today}

\begin{abstract}
We characterize a novel Josephson parametric amplifier based on a flux-tunable quarter-wavelength resonator. The fundamental resonance frequency is \mbox{$\sim$1\,GHz}, but we use higher modes of the resonator for our measurements. An on-chip tuning line allows for magnetic flux pumping of the amplifier. We investigate and compare degenerate parametric amplification, involving a single mode, and nondegenerate parametric amplification, using a pair of modes. We show that we reach quantum-limited noise performance in both cases, and we show that the added noise can be less than 0.5 added photons in the case of low gain.
\end{abstract}


\maketitle 

\section{Introduction}

Due to the rapid advances in circuit quantum electrodynamics (cQED), a promising architecture for quantum information processing, there has been an increased interest in quantum-limited microwave amplifiers in recent years.\cite{Caves1982,Clerk2010} Amplifiers approaching this limit of minimally-added noise have been developed in a number of different superconducting technologies, such as DC-SQUID (superconducting quantum interference device) amplifiers,\cite{Muck2001,Spietz2010} traveling-wave parametric amplifiers,\cite{HoEom2012,OBrien2014,White2015} and resonator-based parametric amplifiers.\cite{Yamamoto2008,Tholen2007,Tholen2009} In particular, systems based on Josephson junctions have been very successful and have found widespread use. For instance, Josephson parametric amplifiers have been used for the generation and measurement of nonclassical states of light,\cite{Mallet2011} quantum-limited measurement of nanomechanical oscillators,\cite{Teufel2009} readout schemes for superconducting qubits,\cite{Siddiqi2004,Abdo2011} and quantum feedback.\cite{Vijay2012}

Quantum-limited performance in Josephson parametric amplifiers (JPA) has been reached in a number of configurations, all based on the modulation of the nonlinear inductance of a number of Josephson junctions. Often the Josephson junctions are configured in a SQUID geometry, or in an array of multiple SQUIDs. The junctions can be embedded in a resonant environment consisting of either a distributed circuit made up of one\cite{Yurke1989,Castellanos-Beltran2007,Yamamoto2008,Eichler2011,Yaakobi2013} or multiple\cite{Bergeal2010a,Roch2012} cavities, a lumped-element circuit,\cite{Hatridge2011,Mutus2013,Zhou2014} or a combination of both.\cite{Mutus2014}

The Josephson inductance can be modulated in two different ways. The first option is by current pumping,\cite{Yurke1989,Hatridge2011} where a strong tone at the signal input port modulates the superconducting phase difference across the junctions. The second way is to flux pump the Josephson inductance, using an on-chip fast tuning line to modulate the flux through the SQUID.\cite{Yamamoto2008,Sandberg2008a,Sundqvist2013,Sundqvist2014} The current pumping case has been explored extensively in the context of parametric amplification by a number of groups. The full nonlinear dynamics of the flux-pumped system has been studied both theoretically and experimentally.\cite{Wilson2010,Dykman1998,Wustmann2013} It has also been shown that flux-pumping can lead to very broadband parametric downconversion even in the absence of a cavity.\cite{Wilson2011}

In this work, we present measurements of a JPA based on a superconducting coplanar waveguide (CPW) resonator. Usually, the fundamental mode of the system is used for parametric amplification. The novelty in our device is that it is designed to have a relatively low fundamental frequency (lower than the cutoff of our measurement band). This allows us access to multiple higher modes within our measurement band.  For a linear resonator, these modes would be equally spaced, but the nonlinearity of the SQUID introduces an anharmonicity in the mode spectrum. Using the higher modes we explore both degenerate, phase-insensitive parametric amplification (single-mode pumping scheme), where the pump is resonant with twice the mode frequency, and nondegenerate parametric amplification (multimode pumping scheme), where the pump is resonant with the sum of the resonance frequencies of two different modes. We make a comparison of these different operation schemes and study the gain, added noise, gain-bandwidth product, and saturation power. We have calibrated our measurement setup with a shot-noise tunnel junction\cite{Spietz2003,Spietz2006}(SNTJ) and demonstrate that the amplifier reaches quantum-limited performance in both operation schemes. 

Compared to single-mode operation, the multimode pumping scheme also gives us access to the full instantaneous bandwidth of one of the modes, without having to worry about interference with the idler, as the idler occurs at a well-separated frequency. Beyond parametric amplification, the multimode pumping has other potential applications.  For instance, when amplifying an input vacuum state, the output photons in the two modes should exhibit two-mode squeezing, a form of continuous-variable entanglement.\cite{Johansson2013} This entanglement generation, together with previous results showing coherent mode conversion in a similar setup,\cite{Zakka-Bajjani2011} is a promising candidate for continuous-variable quantum computing using cluster states.\cite{Bruschi2013}

\section{Sample design and Measurement Setup}
\begin{figure}
\includegraphics{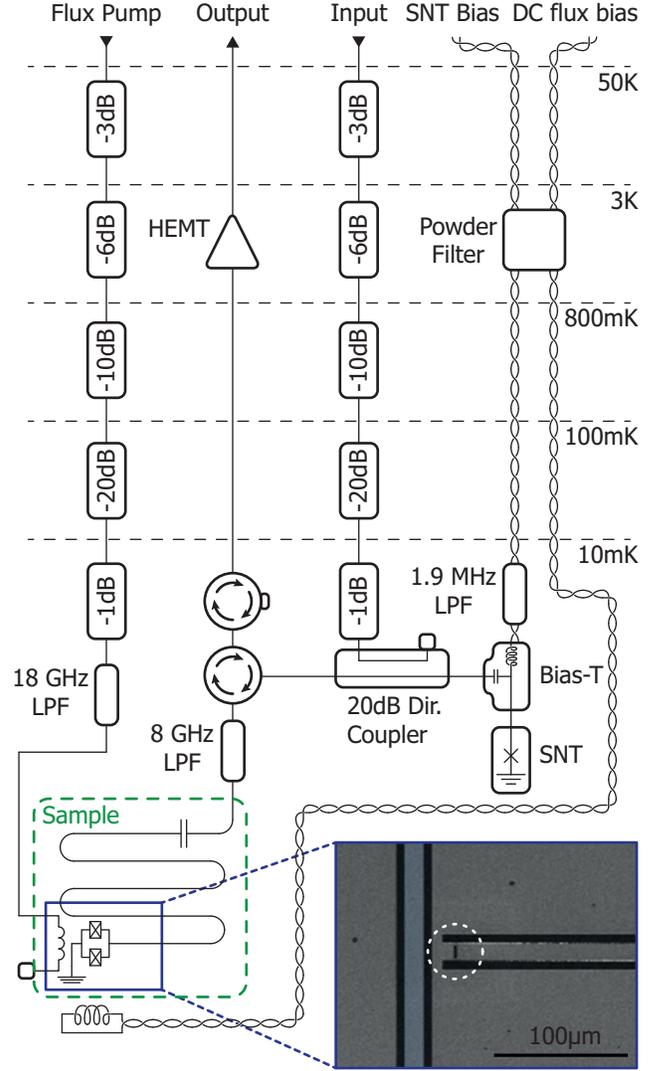}%
\caption{\label{FigSetup}The parametric amplifier (green dashed rectangle) is installed in a microwave reflectometry setup. The input signal is attenuated by \mbox{40\,dB}, distributed over the different temperature stages. At the mixing chamber it is combined with the output of the shot-noise tunnel junction (SNTJ) in a \mbox{20\,dB} directional coupler. The reflected output tone is separated from the input tone by means of two circulators, before being amplified with a cryogenic low-noise HEMT amplifier, installed on the \mbox{3\,K} stage. The DC magnetic flux bias is applied by means of an external coil, while an on-chip inductively-coupled tuning line, attenuated by \mbox{40\,dB}, is used for AC flux pumping. The micrograph shows the AC flux line (blue false color) and the SQUID (white dashed circle). Calibration of the setup is performed with an SNTJ which is installed on the mixing chamber and DC biased through a bias-T.}%
\end{figure}

Our parametric amplifier consists of a superconducting quarter-wavelength coplanar waveguide resonator terminated to ground by means of a SQUID. We fabricate the circuit using electron-beam lithography and standard two-angle evaporation of aluminum. We installed the amplifier in a microwave reflectometry setup (see Fig.~\ref{FigSetup}) and cooled it down in a dilution refrigerator with a base temperature of \mbox{10\,mK}. The inductance of the SQUID, and thus the resonance frequency of the circuit, can be tuned by applying a magnetic flux through the SQUID loop.\cite{Sandberg2008} There is an external coil mounted on the sample box which allows us to apply a DC flux, $\Phi_{\mathrm{DC}}$, and an on-chip CPW line for applying an AC flux, $\Phi_{\mathrm{AC}}$. To calibrate the measurement setup we also installed an SNTJ at the mixing chamber. The SNTJ is DC biased with a bias-T, and the RF noise is combined with the input signal by means of a directional coupler.

\begin{figure}
\includegraphics{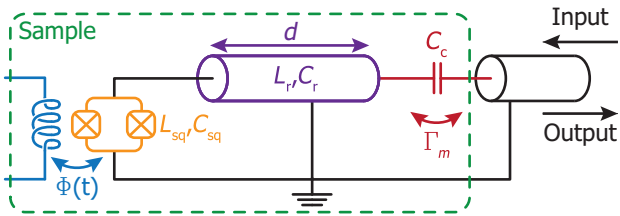}%
\caption{\label{FigTheory}Schematic overview of the resonator. The resonator is marked in purple, with length $d$ and total inductance and capacitance of $L_{\mathrm{r}}$ and $C_{\mathrm{r}}$, respectively. The resonator is connected to ground by means of a SQUID (orange), which is threaded by an external flux $\Phi (t)$ using an on-chip tuning line (blue). The SQUID inductance is $L_{\mathrm{sq}}$ and its capacitance is $C_{\mathrm{sq}}$. The resonator is coupled to the incoming transmission line by a coupling capacitor, $C_{\mathrm{c}}$ (red), giving a mode-dependent coupling rate $\Gamma_m$ for mode $m$.}%
\end{figure}

\begin{figure}
\includegraphics{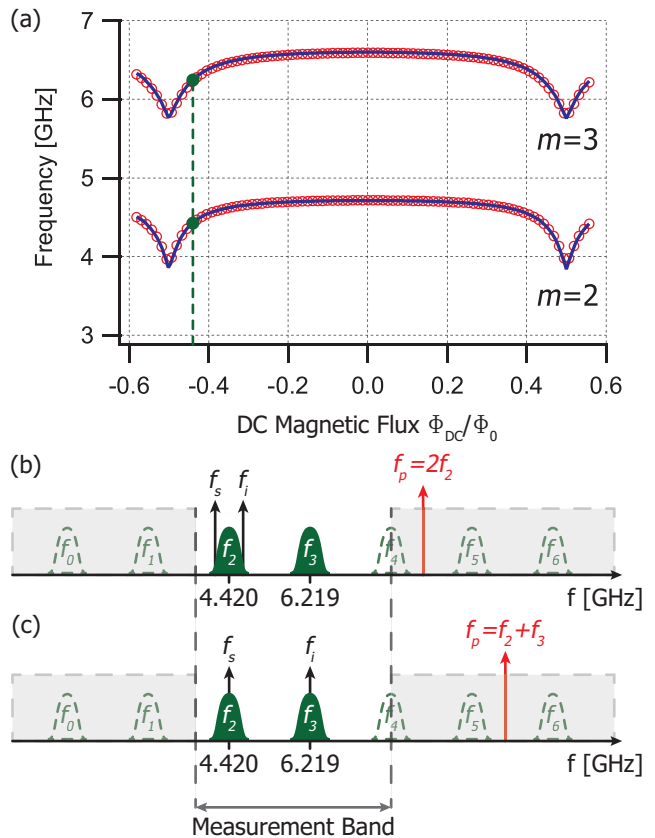}%
\caption{\label{FigTuningPumpScheme}(a) Measured resonance frequencies of the different modes ($m=2,3$) as a function of DC magnetic flux $\Phi_{\mathrm{DC}}$ (red circles). At zero DC flux bias, the resonance frequencies of modes 2 and 3 are, respectively, \mbox{4.713\,GHz} and \mbox{6.588\,GHz}. The blue line a fit to the mode resonance frequencies $f_{m}(\Phi_{\mathrm{DC}})$. The resonances were fit simultaneously by numerically solving the characteristic equation~(\ref{eqnDispRel}). Our Josephson parametric amplifier is operated by magnetic flux-pumping. This is achieved by applying an AC tone to the on-chip flux line, while keeping the DC flux constant at a fixed nonzero value. Our measurements are performed around a DC flux of $\sim-0.44\Phi_{\mathrm{0}}$, denoted by the green dashed line. The unpumped mode frequencies are \mbox{4.420\,GHz} and \mbox{6.219\,GHz} for modes 2 and 3, respectively. (b) When we apply the pump tone at $f_{\mathrm{p}}=2\times f_{2}$ and the signal tone $f_{\mathrm{s}}$ falls within the linewidth of mode 2. The idler tone $f_{\mathrm{i}}$ is created such that $f_{\mathrm{s}}+f_{\mathrm{i}}=f_{\mathrm{p}}$, \textit{i.e.} symmetric around $f_{\mathrm{p}}/2$. (c) When we apply the pump tone at $f_{\mathrm{p}}=f_{2}+f_{3}$ and the signal $f_{\mathrm{s}}$ falls withing the linewidth of mode 2. The idler is now again created symmetrically around $f_{\mathrm{p}}/2$ with respect to the signal and ends thus up in mode $3$. Note that the linewidths in panels (b) and (c) are not to scale. The grey marked regions are outside of the available measurement band of our setup (\mbox{4-8\,GHz}).}%
\end{figure}

We designed the device with a fundamental resonance frequency $f_{m=0}(\Phi_{\mathrm{DC}}=0)$ close to \mbox{1\,GHz}, such that we have access to several higher harmonics in the available measurement band (\mbox{4-8\,GHz}, see Fig.~\ref{FigTuningPumpScheme}). The flux dependence of the mode frequencies ($m=0,1,2,\dots$) is described by the following characteristic equation,\cite{Wallquist2006,Wustmann2013}
\begin{equation}\label{eqnDispRel}
\frac{\pi f_{m}}{2f_{\mathrm{r}}}\tan\left(\frac{\pi f_{m}}{2f_{\mathrm{r}}}\right)=\left|\cos\left(\frac{\pi \Phi_{\mathrm{DC}}}{\Phi_{\mathrm{0}}}\right)\right|\frac{L_{\mathrm{r}}}{L_{\mathrm{sq}}}-\frac{\pi f_{m}}{2f_{\mathrm{r}}}\frac{C_{\mathrm{sq}}}{C_{\mathrm{r}}},
\end{equation}
where $L_{\mathrm{sq}}$ and $C_{\mathrm{sq}}$ denote the SQUID inductance (at $\Phi_{\mathrm{DC}}=0$) and capacitance, respectively (see Fig.~\ref{FigTheory}). $L_{\mathrm{r}}$ and $C_{\mathrm{r}}$ are the inductance and capacitance of the CPW resonator. $f_{\mathrm{r}}$ is the fundamental mode of the bare resonator, neglecting the SQUID (\textit{i.e.} when the CPW is connected directly to ground) and $\Phi_{\mathrm{0}}=h/(2e)$ is the magnetic flux quantum. We used the higher modes $m=2,3$ for our measurements. The resonance frequencies are then found at $f_{m}(0) \approx (2m+1)\times f_{0}(0)$ and can be tuned down to $f_{m}(\Phi_{\mathrm{0}}/2) \approx (2m)\times f_{0}(0)$ (see Fig.~\ref{FigTuningPumpScheme}). In a separate uncalibrated wide-band setup, we have also made measurements using mode 4 (results not shown).

From a normal-state resistance measurement of the SQUID we estimate its inductance $L_{\mathrm{sq}}$ to be approximately $\mbox{200\,pH}$ with a critical current of \mbox{1.65\,$\mu$A}. Simultaneous numerical fitting of the different modes ($m=2,3$) to the characteristic equation, eq.~(\ref{eqnDispRel}), results in a SQUID inductance participation ratio $\gamma=L_{\mathrm{sq}}/L_{\mathrm{r}}\approx1.76\%$. The SQUID capacitance participation ratio $C_{\mathrm{sq}}/C_{\mathrm{r}}$ was found to be much smaller than $\gamma$, and can therefore be neglected in the fits. The fitting also results in a bare resonator frequency $f_{\mathrm{r}}=959\mbox{\,MHz}$. The flux-dependent mode frequencies $f_{m}(\Phi_{\mathrm{DC}})$ for this sample, as well as the fitting results, are shown in Fig.~\ref{FigTuningPumpScheme}(a). 

An interdigitated coupling capacitor, $C_{\mathrm{c}}$, with a capacitance of \mbox{53\,fF} is used to couple the resonator to the \mbox{50\,$\Omega$} measurement line, see Fig.~\ref{FigTheory}. The device is strongly overcoupled, meaning that the internal losses are significantly smaller than the external ones due to the output coupling ($Q_{int}\approx3750$). The resulting loaded quality factor ($Q_{m}$) is therefore limited by the external quality factor, which depends on the mode number. At zero flux, $Q_{m}$ is about 500 for mode 2 and 460 for mode 3.

\section{Theory}
The setup as described above allows for both current- and flux pumping, the latter of which is described below. Applying an AC flux-pumping tone can be done in two different ways: degenerate pumping, where the pump frequency is close to twice the mode frequency ($f_{\mathrm{p}}\approx2\times f_{m}$), and nondegenerate pumping, where the pump frequency is resonant with the sum of two different modes ($f_{\mathrm{p}}\approx f_{m}+f_{n}$). In the degenerate case any signal that falls within the linewidth of the pumped mode of the resonator gets amplified, and the idler is generated in the same mode. In the special narrrow-band case when the the signal frequency is exactly half the pump frequency, signal and idler frequencies are equal ($f_{\mathrm{s}}=f_{\mathrm{i}}$). This is phase-sensitive degenerate amplification, where signal and idler interfere to provide quadrature-dependent gain. In this operation scheme, the amplifier can, in principle, operate without adding noise.\cite{Caves1982}

However, when there is a small offset between $f_{\mathrm{p}}/2$ and $f_{\mathrm{s}}$, \textit{e.g.} where $\left|f_{\mathrm{s}}-f_{\mathrm{p}}/2\right|$ is smaller than the linewidth of the pumped mode, the idler is generated symmetrically around $f_{\mathrm{p}}/2$ with respect to the signal, but still falls within the same mode ($f_{\mathrm{s}}\approx f_{\mathrm{i}}\approx f_{m}$). This is phase-insensitive degenerate amplification and is referred to as the single-mode pumping scheme (see Fig.~\ref{FigTuningPumpScheme}(b)). The gain is now quadrature independent and the minimal noise added by the amplifier is given by the standard quantum limit, \textit{i.e.} equal to \mbox{0.5 photons}, for an amplifier with infinite gain.\cite{Shimoda1957,Haus1962,Caves1982}

Another option is to apply an AC flux-pumping tone which is resonant with the sum of two different modes ($f_{\mathrm{p}}\approx f_{m}+f_{n}$). In this case any signal falling within the linewidth of one of the two different modes is amplified, generating the idler symmetrically around $f_{\mathrm{p}}/2$ with respect to the signal. The idler is thus generated in the other mode, ($f_{\mathrm{s}}\approx f_{m}$ and $f_{\mathrm{i}}\approx f_{n}$).  This is nondegenerate amplification and we refer to this as the multimode pumping scheme (see Fig.~\ref{FigTuningPumpScheme}(c)). As this pumping scheme is inherently phase insensitive ($f_{\mathrm{s}}\neq f_{\mathrm{i}}$ and signal and idler can thus never interfere), the minimal noise added by the amplifier is also here the standard quantum limit.\cite{Shimoda1957,Haus1962,Caves1982}

The single-mode pumping scheme is theoretically described in the derivation of Wustmann \textit{et al.},\cite{Wustmann2013} where the behavior of the circuit is analyzed in the case of a harmonic flux drive around a DC flux bias, $\Phi(t) = \Phi_{\mathrm{DC}} + \Phi_{\mathrm{AC}} \cos(2\pi f_{\mathrm{p}} t)$. We can extend this description and show that the full flux-pumped circuit behaves as a set of harmonic oscillators, one for each mode, that are coupled through a time-dependent, nonlinear potential. The system exhibits a wide variety of interesting dynamical features. In particular, it exhibits a number of characteristic ``resonances'' when the parametric pump frequency $f_{\mathrm{p}}$ is equal to the sum or difference of two mode frequencies, \textit{i.e.}, when $f_{\mathrm{p}} \approx f_m \pm f_n$. In the case where $f_{\mathrm{p}} \approx f_m - f_n$, we expect to see intermode conversion with a ``beam splitter'' type interaction.\cite{Zakka-Bajjani2011} When $f_{\mathrm{p}} \approx f_m + f_n$, we expect to see parametric amplification and oscillations.

Following the method outlined in Wustmann \textit{et al.}\cite{Wustmann2013} and extending it to the multimode case, we derive the equation for the multimode gain, $G_{mn}$,  
\begin{eqnarray}\label{Gain}
&G_{mn} = \delta_{mn} + \frac{4\epsilon^2\Gamma_m\Gamma_n}{(\Delta^2 - \Delta_0^2 + \epsilon^2 - \epsilon_{\mathrm{th}}^2)^2 + (\Gamma_m +\Gamma_n)^2(\Delta-\Delta_0)^2}.
\end{eqnarray}
Here $\delta_{mn}$ is the Kronecker delta, $\epsilon$ is the effective pump strength,
\begin{eqnarray}\label{PumpAmp}
&\epsilon = \frac{\pi\Phi_{\mathrm{AC}}}{\Phi_{0}}\frac{\sin(\pi\frac{\Phi_{\mathrm{DC}}}{\Phi_{0}})}{2\gamma}\frac{\sqrt{2\pi f_m}\cos(k_m d)}{\sqrt{M_m}(k_m d)}\frac{\sqrt{2\pi f_n}\cos(k_n d)}{\sqrt{M_n}(k_n d)},
\end{eqnarray}
$\Gamma_m=\pi f_{m}/Q_m$ is the coupling rate of mode $m$ to the transmission line (see Fig.~\ref{FigTheory}), it is given by equation,\cite{Wustmann2013}
\begin{eqnarray}
\Gamma_m=2\pi f_m\left(\frac{C_c}{C_r}\right)^{2}\frac{k_m d}{M_m}
\end{eqnarray}
where $C_c$ is the coupling capacitance and $C_r$ is the cavity capacitance.
In these equations, quantity $M_m$ is given by equation,\cite{Wustmann2013}
\begin{equation}
M_m = 1+\frac{\sin\left(2 k_m d\right)}{2 k_m d}+\frac{2 C_{\mathrm{sq}}}{C_{\mathrm{r}}}\cos^2\left(k_m d\right).
\end{equation}
$\gamma$ is the SQUID inductance participation ratio, $d$ is the cavity length, $k_m$ is the wavenumber, 
for our low $\gamma$, $k_m d \approx\pi(1/2+m)$. $\epsilon_{\mathrm{th}}^2 = \Gamma_m\Gamma_n + \delta^2 -\Delta_0^2$ is the parametric oscillation threshold. The pump frequency detuning is denoted by $\delta = 2\pi\left( f_{\mathrm{p}}/2 - (f_m + f_n)/2\right)$. Finally we also introduced an asymmetry parameter $\Delta_0$, taking into account the different coupling rates $\Gamma_m$ and $\Gamma_n$ for the different modes:%
\begin{equation}
\Delta_0 = \delta \;{\Gamma_n - \Gamma_m\over \Gamma_n + \Gamma_m}.
\end{equation}

In eq.~(\ref{Gain}), the signal detuning $\Delta$ is referenced to the mode frequency plus the pump detuning as follows:  
\begin{equation}
\Delta = 2\pi f_s -(2\pi f_m+\delta).
\end{equation}
Note that the idler is found at a frequency $2\pi f_i=\left(2\pi f_n+\delta\right)-\Delta$, such that $f_s$ and $f_i$ are located symmetrically around $f_{\mathrm{p}}/2$.

In general, the gain peaks are not Lorentzian. Moreover, for a nonzero pump detuning $\delta$, the gain exhibits two resonance peaks in the parameter region far from the parametric oscillation threshold (\textit{i.e.} when $\epsilon<<\epsilon_{\mathrm{th}}$), similar to the degenerate case.\cite{Wustmann2013} These peaks are, however, asymmetric in position and height. While approaching the threshold, the peaks merge into a single peak, which is shifted from $f_m+\delta$ by $\Delta_0$ near the threshold. For $\delta=0$ we always have a single peak when pumping below the threshold.

For pump strengths $\epsilon$ close to the threshold, $\epsilon_{\mathrm{th}}$, the single resonance gain peak can be approximated as a Lorentzian,
\begin{eqnarray}\label{GainLOR}
&G_{mn}(\Delta)\approx \\
&\delta _{mn} + {4\epsilon_{\mathrm{th}}^2 \chi \Gamma_m\Gamma_n \over 
[\Delta - \Delta_0 + 2\chi\Delta_0(\epsilon^2-\epsilon_{\mathrm{th}}^2)]^2 + \chi^2(\Gamma_m +\Gamma_n)^2
(\epsilon^2 - \epsilon_{\mathrm{th}}^2)^2}\nonumber
\,,
\end{eqnarray}
with $\chi = [4\Delta_0^2 + (\Gamma_m +\Gamma_n)^2]^{-1}$.
In this approximation we can unambiguously define the bandwidth: 
\begin{equation}
{\rm BW} =  2\chi(\Gamma_m +\Gamma_n) (\epsilon^2 - \epsilon_{\mathrm{th}}^2).
\end{equation}
The amplitude gain is then $\sqrt G = 2\epsilon\sqrt{\chi \Gamma_m\Gamma_n }/ ({\rm BW}/2)$. We can make a zeroth order estimate of the gain-bandwidth product by replacing $\epsilon$ with $\epsilon_{\mathrm{th}}$ which gives:
\begin{equation}\label{GBWPMulti}
\sqrt G \cdot {\rm BW} = 4\epsilon_{\mathrm{th}}\sqrt{\chi \Gamma_m\Gamma_n }.
\end{equation}
In the single-mode case, with $\Gamma_n=\Gamma_m =\Gamma$ and $\chi = 1/4\Gamma^2$, this becomes:
\begin{equation}\label{GBWPSingle}
\sqrt G \cdot {\rm BW} = 2\epsilon_{\mathrm{th}} = 2\Gamma \,.
\end{equation}

\section{Results}
\subsection{Single-mode pumping}
We first analyze the single-mode pumping scheme for mode $m=2$. The mode frequency, $f_2=4.420\mbox{\,GHz}$, is determined by the applied DC flux, which in our measurements was $\Phi_{\mathrm{DC}}\sim-0.44\Phi_{\mathrm{0}}$. When we flux pump the SQUID in order to study parametric amplification, there is a nonlinear pump-induced frequency shift of the resonance to a slightly lower frequency.\cite{Wustmann2013,Krantz2013} As this frequency shift depends on the pump strength, we have to sweep the pump over an appropriate frequency range in order to find the resonance for each pump power.

\begin{figure}
\includegraphics{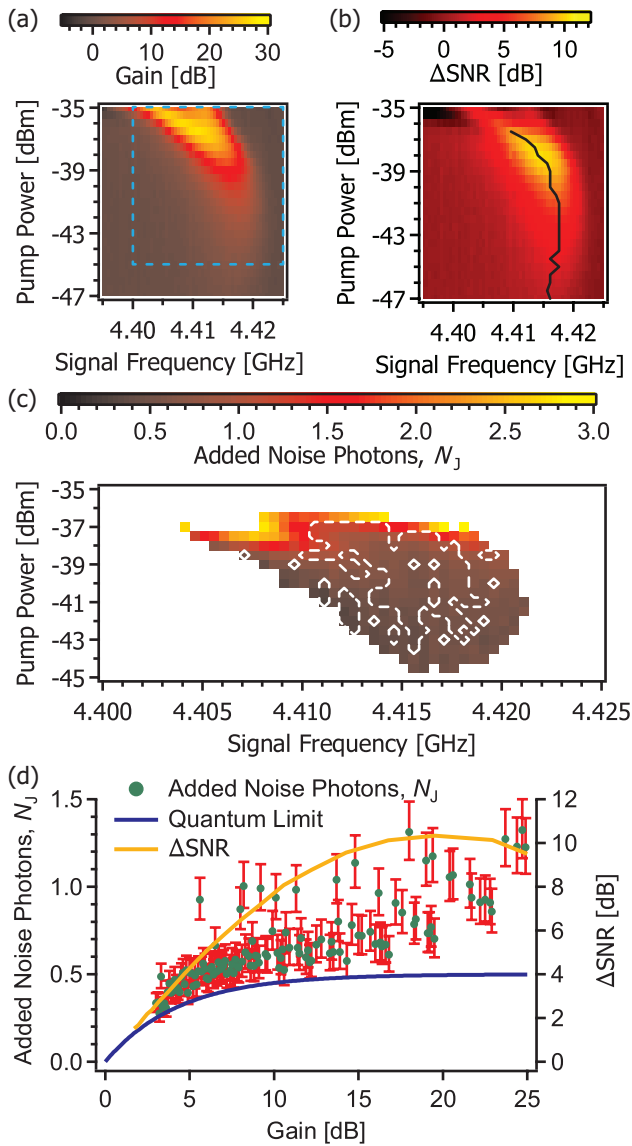}%
\caption{\label{FigSingleGainNoise}Single-mode pumping. (a) Power gain, (b) improvement in SNR ($\Delta \mathrm{SNR}$) and (c) number of added noise photons ($N_{\mathrm{J}}$) observed at $f_{\mathrm{s}}=f_{\mathrm{p}}/2+100\mbox{\,kHz}$ as a function of pump power, $P_{\mathrm{p}}$, and signal frequency, $f_{\mathrm{s}}$. The range of panel (c) is marked by the blue dashed rectangle in panel (a). The optimal point of operation, where we get the highest SNR improvement, is at $f_{\mathrm{s}}=4.4141\mbox{\,GHz}$ and $P_{\mathrm{p}}=-38\mbox{\,dBm}$, and shows $\Delta \mathrm{SNR}=\mbox{10.5\,dB}$. Whenever the gain exceeded \mbox{3\,dB}, we extracted the noise added by the parametric amplifier. The maximum $\Delta \mathrm{SNR}$ for each pump power is marked with the black line in panel (b). Panel (d) shows the minimum added noise as a function of gain (in $0.1\mbox{\,dB}$ wide bins). The error bars reflect the uncertainty in the insertion loss of the components installed between the SNTJ and the parametric amplifier (\mbox{1.75$\pm$\,0.4dB}). The blue line marks the quantum limit as a function of gain (eq.~(\ref{eqQL})). All the points of added noise in panel (d) fall within the white dashed contours in panel (c). The orange line shows the $\Delta \mathrm{SNR}$ as a function of gain taken along the black line in panel (b).}%
\end{figure}

To map out the region in which we observe gain, we scanned the pump both in power and in frequency. A small signal tone was applied with an offset of $\Delta=100\mbox{\,kHz}$ compared to half the pump frequency, $f_{\mathrm{s}}=f_{\mathrm{p}}/2+100\mbox{\,kHz}$, to operate the device nondegenerately (such that the gain does not depend on the pump phase). We extracted the gain by comparing the reflected signal power with the pump on and off (see Fig.~\ref{FigSingleGainNoise}). We also measured the increase of the noise floor at the signal frequency with the pump on compared to off. The gain and the increase in the noise floor allow us to calculate the improvement in signal-to-noise ratio ($\Delta \mathrm{SNR}$, see Fig.~\ref{FigSingleGainNoise}(b)) provided by the parametric amplifier, $\Delta \mathrm{SNR}=\mathrm{SNR}_{\mathrm{H,J}}/\mathrm{SNR}_{\mathrm{H}}$. $\mathrm{SNR}_{\mathrm{H}}$ denotes the SNR with only the HEMT amplifier on, and $\mathrm{SNR}_{\mathrm{H,J}}$ is the SNR when we also turn on the parametric amplifier. Note that this ratio is taken in linear units. The amplifier performance is optimal when $\Delta \mathrm{SNR}$ is maximized. We find a maximal $\Delta \mathrm{SNR}$ of \mbox{10.5\,dB} for a pump strength of $-38\mbox{\,dBm}$ and a pump frequency of $8.828\mbox{\,GHz}$.

The noise performance of amplifiers is often quantified by the amount of noise that they add, referred to their input. There exists, however, a quantum limit which puts a lower limit on the amount of noise added. For a nondegenerate amplifier (which amplifies both quadratures equally), the quantum limit of added noise ($N_{\mathrm{QL}}$) depends on the power gain of the amplifier ($G$) in the following way:\cite{Shimoda1957,Haus1962,Caves1982}
\begin{equation}\label{eqQL}
N_{\mathrm{QL}} = \frac{1}{2}\left|1-\frac{1}{G}\right|.
\end{equation}
Note that the quantum limit tends to 0.5 added photons when the gain of the amplifier goes to infinity. However, for finite gain the added number of photons can become less than 0.5 photons.

To calculate the amount of noise added by our parametric amplifier (as referred to its input), we need to calibrate the noise of the measurement chain without the parametric amplifier. For this purpose we installed an SNTJ at the mixing chamber of our cryostat. The SNTJ acts as a controllable noise source, allowing us to calibrate the total noise as referred to the SNTJ ($N_{\mathrm{SNTJ}}$), over the frequency range of interest. During the calibration, the pump is turned off and the resonator is tuned away from the band of interest. Then, the measurement chain was found to add $\sim24.5\mbox{\,photons}$ for a total noise $N_{\mathrm{SNTJ}}\sim25\mbox{\,photons}$. In order to calculate the added noise by the parametric amplifier itself, we convert this number to the input of the parametric amplifier. To do this we need to take the insertion loss of the components between the SNTJ and the parametric amplifier into account (a bias-T, directional coupler, circulator, and 8\,GHz low-pass filter). Assuming that these components are at the same temperature as the SNTJ, this is done as follows:
\begin{equation}\label{eqNH}
N_{\mathrm{SYS}} = \frac{N_{\mathrm{SNTJ}}}{A_{IL}}-N_{\mathrm{in}},
\end{equation}
In eq.~(\ref{eqNH}), $N_{\mathrm{SYS}}$ is the noise of the measurement chain, referred to the input of the parametric amplifier, $A_{\mathrm{IL}}$ is the total insertion loss of the components in the path between the SNTJ and the parametric amplifier and $N_{\mathrm{in}}$ is the number of noise photons of the field in the input line, which is calibrated at the SNTJ to be very close to vacuum noise, 0.5\,photons. With an insertions loss $A_{\mathrm{IL}}=1.75\pm0.4\mbox{\,dB}$, verified in a separate measurement, we obtain \mbox{$N_{\mathrm{SYS}}\sim16.2^{+1.6}_{-1.5}$ photons}, which corresponds to a noise temperature of $\sim3.4\mbox{\,K}$. The factory-measured noise temperature of our HEMT amplifier is $\sim2\mbox{\,K}$. This means that the $N_{\mathrm{SYS}}$ corresponds to a loss of $\sim2\mbox{\,dB}$ between the parametric amplifier and the HEMT, which is reasonable.

As we now know the noise added by our measurement chain, the gain of the parametric amplifier and the improvement in SNR, we can rewrite $\Delta \mathrm{SNR}$ as a function of the amount of noise, $N_{\mathrm{J}}$, added by the parametric amplifier as follows:
\begin{eqnarray}\label{eqDeltaSNR}
\Delta \mathrm{SNR} & = & \frac{\frac{G_{\mathrm{SYS}}\,G_{\mathrm{J}}\,S}{G_{\mathrm{SYS}}\left(N_{\mathrm{SYS}}+G_{\mathrm{J}}\left(N_{\mathrm{J}}+N_{\mathrm{in}}\right)\right)}}{\frac{G_{\mathrm{SYS}}\,S}{G_{\mathrm{SYS}}\left(N_{\mathrm{SYS}}+N_{\mathrm{in}}\right)}}\nonumber\\
& = & \frac{N_{\mathrm{SYS}}+N_{\mathrm{in}}}{\frac{N_{\mathrm{SYS}}}{G_{\mathrm{J}}}+N_{\mathrm{J}}+N_{\mathrm{in}}}.
\end{eqnarray}
Here $N_{\mathrm{J}}$ and $G_{\mathrm{J}}$ are the number of noise photons added by the parametric amplifier and its power gain, respectively. $G_{\mathrm{SYS}}$ is the gain of the measurement chain with the parametric amplifier turned off. We solve eq.~(\ref{eqDeltaSNR}) to express $N_{\mathrm{J}}$ as:
\begin{equation}\label{EqnAddedNoise}
N_{\mathrm{J}} = N_{\mathrm{SYS}}\left(\frac{1}{\Delta \mathrm{SNR}}-\frac{1}{G_{\mathrm{J}}}\right)+N_{\mathrm{in}}\left(\frac{1}{\Delta \mathrm{SNR}}-1\right).
\end{equation}

Using eq.~(\ref{EqnAddedNoise}), we can calculated the added number of noise photons, $N_{\mathrm{J}}$ from $G_{\mathrm{J}}$ and $\Delta \mathrm{SNR}$. In Fig.~\ref{FigSingleGainNoise}(c), we show $N_{\mathrm{J}}$ for all points where the gain was larger than \mbox{3\,dB}. To see how close we get to the quantum limit, we present the data in an alternative way in Fig.~\ref{FigSingleGainNoise}(d). We combine Figs.~\ref{FigSingleGainNoise}(a) and \ref{FigSingleGainNoise}(c), in a plot of the added noise versus the gain . However, we retain only the points with the lowest $N_{\mathrm{J}}$ for each bin, which is a 0.1\,dB wide range of gain in Fig.~\ref{FigSingleGainNoise}(d). These points are grouped within the white contours of Fig.~\ref{FigSingleGainNoise}(c). We see that $N_{\mathrm{J}}$ follows the quantum limit nicely, even where the quantum limit is significantly smaller than 0.5\,photons. The error bars reflect the uncertainty in the insertion loss of the components installed between the SNTJ and the parametric amplifier as described above. We also plot the maximum $\Delta \mathrm{SNR}$ for each pump power as a function of the power gain at that point.

\begin{figure}
\includegraphics{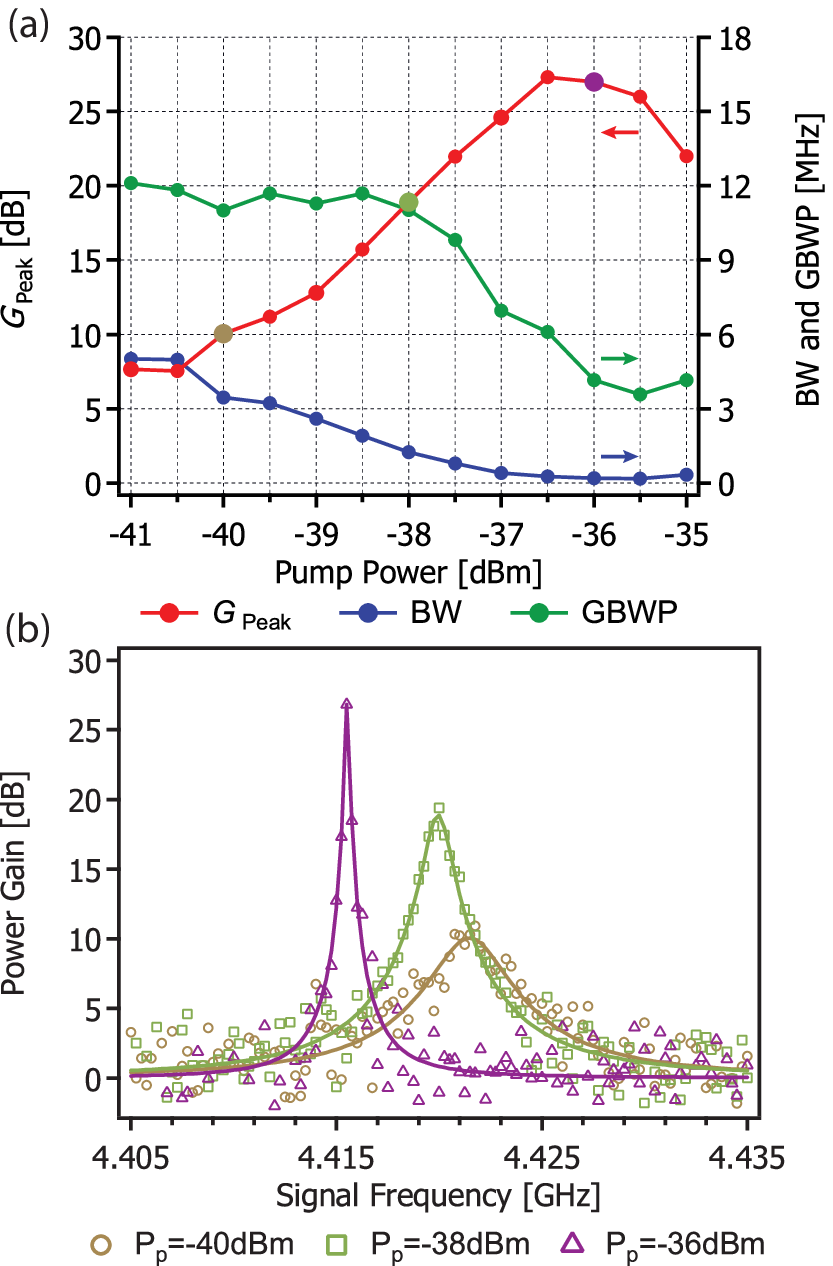}%
\caption{\label{FigSingleGainBW}Single-mode pumping. (a) Peak power gain ($G_{\mathrm{Peak}}$), \mbox{3\,dB} bandwidth (BW), and gain-bandwidth product (GBWP) as a function of pump power (and for the pump frequency where the gain was maximum). The bandwidth was extracted as the FWHM of the gain peak and is shown in blue. The maximum of the gain peak is shown in red. The GBWP is then defined as $\sqrt{G_{\mathrm{Peak}}}\mathrm{BW}$ and was found to have a plateau at \mbox{12\,MHz}. At higher pump powers it starts to drop off. (b) Power gain as a function of signal frequency for several pump powers marked by the colored circles in panel (a). We also present the Lorentzian fit of each gain with the solid lines.}%
\end{figure}

The noise performance is not the only point of interest in an amplifier. The bandwidth and its dependence on the gain are also important. To measure the bandwidth of the amplifier, we need to record the gain as a function of signal frequency for each pump power and frequency. We do this by sweeping the signal tone in a range around $f_{2}$. We record the change in reflection coefficient of the amplifier when the pump was turned from off to on. The bandwidth, BW, is then extracted (by fitting a Lorentzian) as the full width at half maximum of the power gain peak. The peak power gain, $G_{\mathrm{peak}}$, is also extracted (see Fig.~\ref{FigSingleGainBW}(b)). For each pump power, we find the pump frequency which shows the largest gain (where the pump is closest to twice the resonance frequency). Note that the resonance frequency tends to shift down with increasing pump strength as discussed above. We calculate the gain-bandwidth product for the optimal pump frequency as $\mathrm{GBWP}=\sqrt{G_{\mathrm{peak}}}\mathrm{BW}$. $G_{\mathrm{peak}}$, BW and GBWP are shown as a function of pump power in Fig.~\ref{FigSingleGainBW}(a). The GBWP shows a plateau at \mbox{12\,MHz}. When the gain surpasses 20\,dB, the GBWP starts to fall off. The drop in $G_{\mathrm{peak}}$ and GBWP are likely caused by the transition into the parametric oscillation regime.

A last figure of merit is the saturation power of the amplifier. This is defined as the signal power at which the gain is decreased by \mbox{1\,dB}. We extracted the saturation power at a single pump power and frequency, where the gain was \mbox{10.5\,dB}. This value allows for a comparison between the different pumping schemes. In this measurement we fix the pump and sweep the signal frequency again around $f_{\mathrm{p}}/2$. We record the peak gain, $G_{\mathrm{peak}}$ as a function of increasing signal power. The saturation power was found to be \mbox{-133.5\,dBm}.

\subsection{Multimode pumping}
We are not limited to pumping of a single mode, but we can also pump with a frequency $f_{\mathrm{p}}=f_{m}+f_{n}$, with $m\neq n$. In this case, we expect that any signal tone with a frequency close to that of a pumped mode is amplified. In the available measurement bandwidth of our setup, and with the DC flux bias $\Phi_{\mathrm{DC}}=-0.44\Phi_{\mathrm{0}}$, we used two of the available modes, $f_{2}$ and $f_{3}$, with unpumped resonance frequencies \mbox{4.420\,GHz} and \mbox{6.219\,GHz}, respectively.

\begin{figure*}
\includegraphics{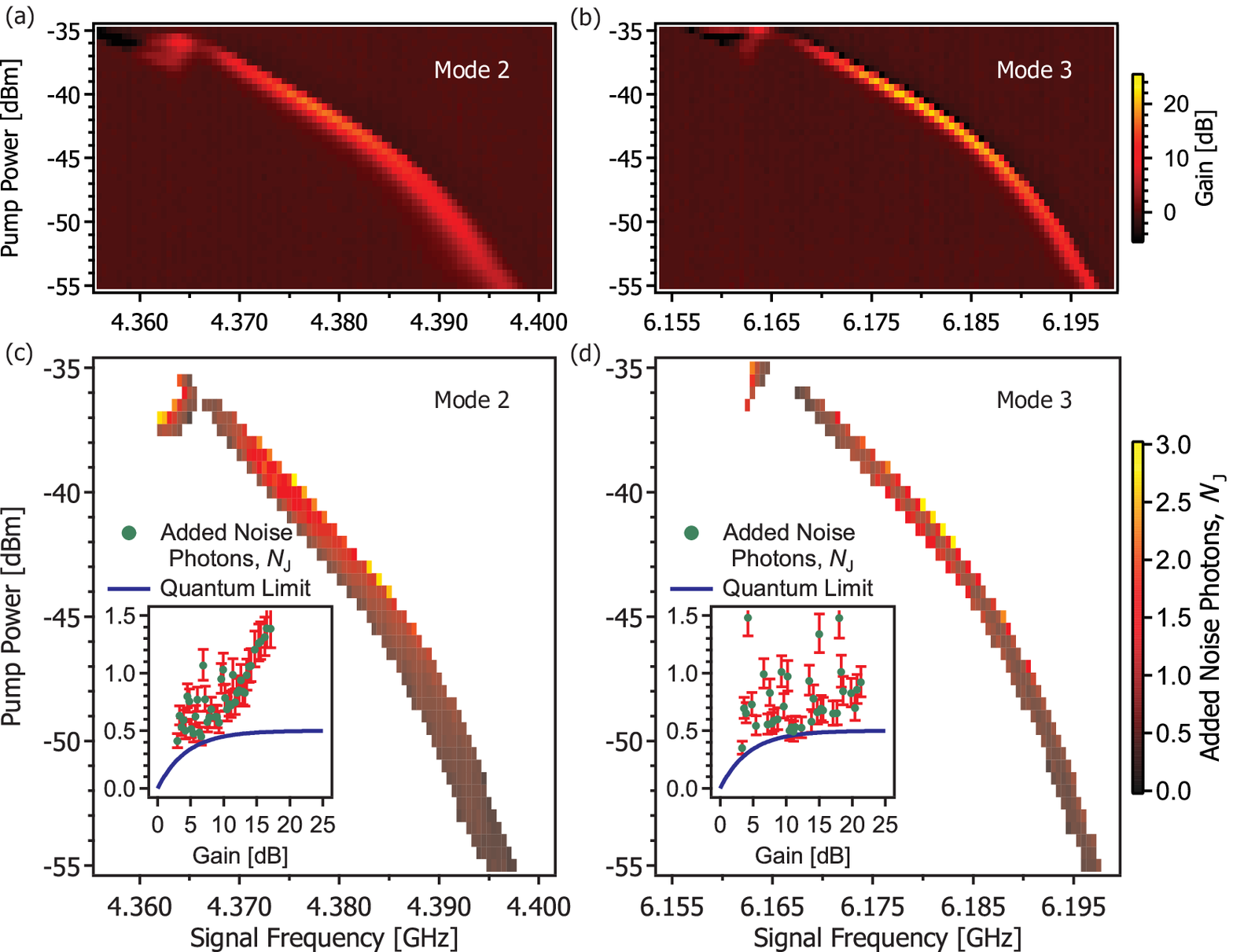}%
\caption{\label{FigMultiGain}Multimode pumping. (a-b) Power gain and (c-d) number of added noise photons ($N_{\mathrm{J}}$) observed at $f_{\mathrm{s}}=f_{\mathrm{p}}/2-899\mbox{\,MHz}$ (mode 2, left column) and at $f_{\mathrm{s}}=f_{\mathrm{p}}/2+899\mbox{\,MHz}$ (mode 3, right column), as a function of pump power $P_{\mathrm{p}}$ and signal frequency $f_{\mathrm{s}}$. Maximum gains of \mbox{17\,dB} and \mbox{22\,dB} are found for modes 2 and 3 respectively. The optimal point of operation shows a maximal $\Delta \mathrm{SNR}$ (not shown) of \mbox{9.5\,dB} and \mbox{10.5\,dB} for modes 2 and 3, respectively. Whenever the gain exceeded \mbox{3\,dB} we extracted the noise added by the parametric amplifier. The insets in panels (c) and (d) show the minimum added noise as a function of gain (in $0.1\mbox{\,dB}$ wide bins). Error bars reflect the uncertainty in the insertion loss of the components installed between the SNTJ and the parametric amplifier (\mbox{1.75$\pm$\,0.4dB} for mode 2 and \mbox{2.25$\pm$\,0.4dB} for mode 3). The blue line marks the quantum limit as a function of gain (eq.~(\ref{eqQL})).}%
\end{figure*}

To map out the region in which we see gain, the pump was scanned both in power and in frequency. A small signal tone was applied at $f_{\mathrm{s}}=f_{\mathrm{p}}/2\pm899\mbox{\,MHz}$. In this way, the signal falls within mode 2 or 3. Note that $\Delta=\pm500\mbox{\,kHz}$ is now larger than in the single-mode pumping scheme. The gain is shown in Fig.~\ref{FigMultiGain}(a-b). We also calculated the improvement in SNR (not shown) in a similar fashion as above, and the optimal point of operation showed a maximal SNR improvement of \mbox{9.5\,dB} and \mbox{10.5\,dB} for modes 2 and 3, respectively.

The noise performance of the multimode pumping scheme is evaluated by calculating $N_{\mathrm{J}}$ in a similar fashion as in the single-mode pumping case (see Fig.~\ref{FigMultiGain}(c-d)). The calibration of the measurement chain gives $N_{\mathrm{SYS}}\sim16.2^{+1.6}_{-1.5}$ photons in mode 2 and $N_{\mathrm{SYS}}\sim13.8^{+1.4}_{-1.3}$ photons in mode 3 (with $A_{\mathrm{IL}}=1.75\pm0.4\mbox{\,dB}$ for mode 2 and $A_{\mathrm{IL}}=2.25\pm0.4\mbox{\,dB}$ for mode 3). As this operation scheme is also nondegenerate, the noise added by the parametric amplifier is bounded by the same quantum limit as the single-mode pumping, $N_{\mathrm{J}}\geq\frac{1}{2}\left|1-\frac{1}{G_{\mathrm{J}}}\right|$. We also present the added noise data in the insets of Fig.~\ref{FigMultiGain}(c-d) in the same way as for the single-mode case.

\begin{figure}
\includegraphics{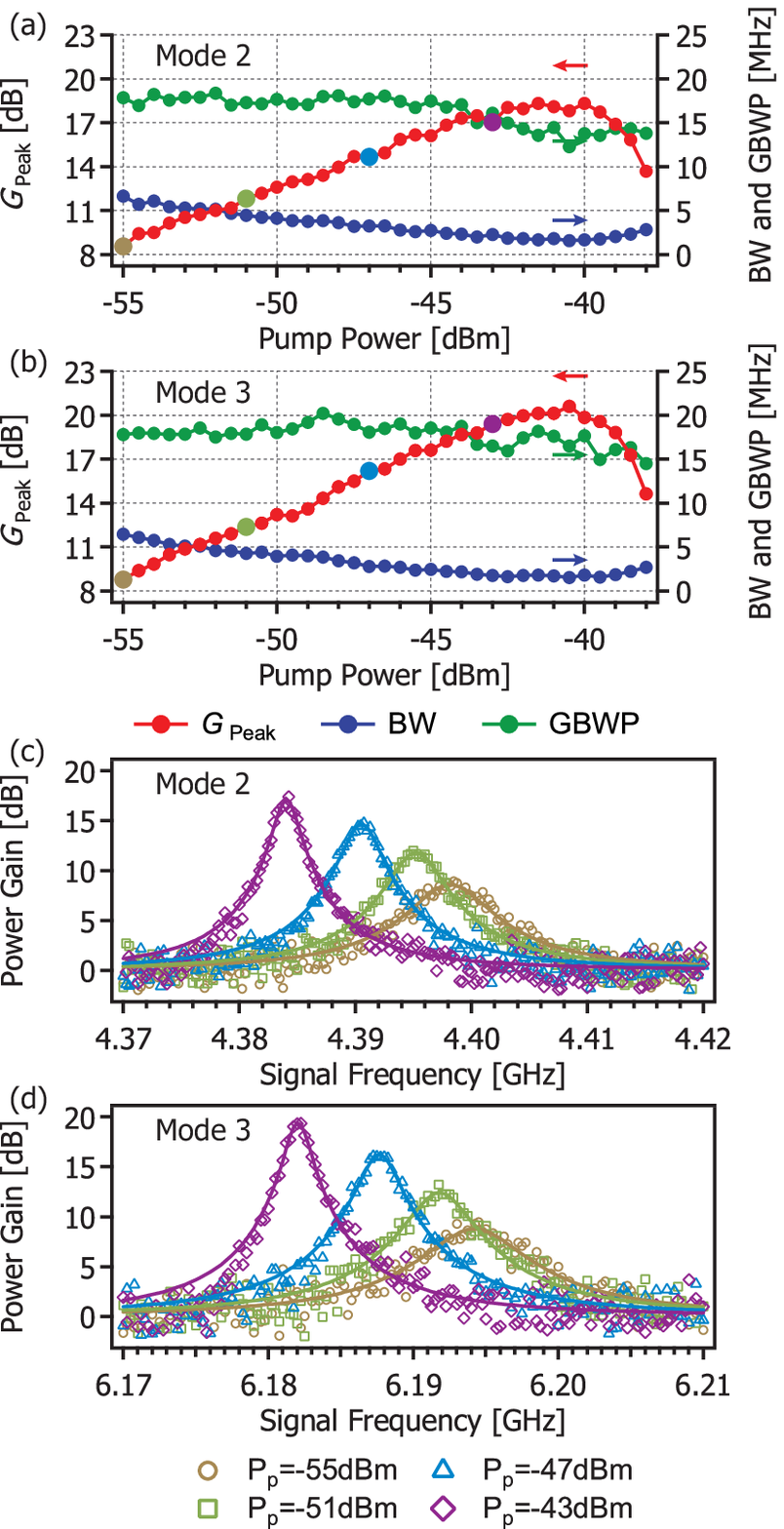}%
\caption{\label{FigMultiGainBW}Multimode pumping. Peak power gain ($G_{\mathrm{Peak}}$), \mbox{3\,dB} bandwidth (BW), and gain-bandwidth product (GBWP) as a function of pump power (and for the pump frequency where the gain was maximum), for mode 2 (a) and 3 (b). The bandwidth was extracted as the FWHM of the gain peak and is shown in blue. The maximum of the gain peak is shown in red. The GBWP is then defined as $\sqrt{G_{\mathrm{Peak}}}\mathrm{BW}$ and was found to have a plateau at \mbox{17.5\,MHz}, for both modes. At higher pump powers it starts to drop off. Power Gain for mode 2 (c) and 3 (d) as a function of signal frequency for several pump powers marked by the colored circles in panels (a-b). We also present the Lorentzian fits of the gain with the solid lines.}%
\end{figure}

Gain, BW, and the GBWP are measured in a similar way as before. The $G_{\mathrm{Peak}}$, BW and GBWP are shown, as a function of pump power in Fig.~\ref{FigMultiGainBW}(a-b). The GBWP shows a plateau at \mbox{17.5\,MHz}, for both modes. When the gain surpasses $\sim\mbox{18\,dB}$, the GBWP starts to drop off. We also show the power gain as a function of $f_{\mathrm{s}}$ for a number of pump powers in Fig.~\ref{FigMultiGainBW}(c-d).

Lastly, we also measured the saturation power of the parametric amplifier with \mbox{11\,dB} of gain, similar to the gain for which the saturation powers in the single-mode case was extracted. The saturation power for modes 2 and 3, were found to be -133\,dBm and -131\,dBm respectively. We present this data and the data from the single-mode pumping scheme in Table~\ref{TablePSAT}.

\begin{table}
\caption{\label{TablePSAT}Gain, saturation power, $P_{\mathrm{Sat}}$, and GBWP for modes 2 and 3 for multimode pumping, and a comparison with the single-mode pumping case.}
\begin{ruledtabular}
\begin{tabular}{ccccc}
Pumping scheme&Mode number&Gain&$P_{\mathrm{Sat}}$&GBWP\\
-&-&[dB]&[dBm]&[MHz]\\
\hline
Single-mode&2&10.5&-133.5&12\\
Multimode&2&11&-133&17.5\\
Multimode&3&11&-131&17.5\\
\end{tabular}
\end{ruledtabular}
\end{table}

\section{Discussion and Summary}
Having characterized the different pumping schemes, we can now make a comparison. Within the investigated pump parameter space, we achieved quantum-limited performance for both single- and multimode pumping. In both cases, the amount of noise added by the parametric amplifier is gain dependent, and drops below 0.5 photons for low gain, in agreement with theory.

The optimal point of operation is where the improvement in SNR, $\Delta\mathrm{SNR}$, is maximized. For the single-mode case the maximal $\Delta\mathrm{SNR}$ was \mbox{10.5\,dB} with $N_{\mathrm{J}}=0.7\mbox{\,photons}$. At this point the total noise of the measurement setup referred to the input of the JPA (including the JPA itself) is \mbox{0.86\,photons}. For the multimode case, the optimum values gave a $\Delta\mathrm{SNR}$ of \mbox{9.5\,dB} ($N_{\mathrm{J}}=0.98\mbox{\,photons}$ and the total added noise of 1.57\,photons) and \mbox{10.5\,dB} ($N_{\mathrm{J}}=0.7\mbox{\,photons}$ and the total added noise of 0.83\,photons) for modes 2 and 3, respectively.

We also analyzed the bandwidth of the amplifier in both schemes. The GBWP was then presented as a function of pump strength, and for the pump frequency which showed maximum gain. By doing this we can assume that the effective pump detuning, taking into account pump-induced frequency shift, is close to zero. Using equations (\ref{GBWPMulti}) and (\ref{GBWPSingle}) to calculate the theoretical value of the GBWP we expect \mbox{13.8\,MHz} and \mbox{19.4\,MHz} for the single- and multimode pumping schemes respectively. In the experiment, the GBWP as a function of pump strength is indeed fairly constant at a level of \mbox{12\,MHz} and \mbox{17.5\,MHz} for the single- and multimode pumping scheme, respectively. These values are relatively close to the theoretically expected values. For gain values larger than 20\,dB, the GBWP starts to fall off. This likely indicates that we have crossed the threshold for parametric oscillations. The drop in GBWP is more significant for the single-mode pumping scheme compared to the multimode pumping scheme.

The multimode pumping scheme has similarities to previous work with a so called Josephson parametric converter.\cite{Bergeal2010,Bergeal2010a,Abdo2011,Roch2012} That device separates the signal and idler modes at two different frequencies into two physically distinct cavities which are then connected by a network of Josephson junctions that requires multilayer fabrication with crossover wiring. The design aims to eliminate the higher-order nonlinearities from the Josephson junctions. However, separation of the signal and idler in different cavities is not always necessary. Our design is less intricate and also allows us to separate signal and idler over different modes of the same resonator.

In summary, we measured a novel Josephson parametric amplifier where we compared two different pumping schemes. The amplifier closely approached quantum-limited noise performance for both the single-mode pumping scheme and the multimode pumping scheme. In accordance with theory, we also observed that the added noise can be less than 0.5 photons for relatively low gain. The multimode pumping scheme, where we pump different pairs of modes of the same resonator, is a novel way of generating parametric amplification. Note that, in this case, we can achieve amplification in frequency bands which are separated by several GHz. In contrast to the single-mode pumping schemes, we can use the whole bandwidth of the amplifier, as the idler occurs at a well-separated frequency.
\begin{acknowledgments}
The authors would like to thank J. Aumentado for providing the SNTJ, for fruitful discussions and useful comments. We acknowledge support from the Swedish Research Council, the Wallenberg foundation, the Swedish Foundation for International Cooperation in Research and Higher Education, the University of Waterloo, NSERC of Canada, and the EU through the ERC, the Marie Curie Actions CIG and the projects ScaleQIT and PROMISCE.
\end{acknowledgments}

\bibliography{References}

\end{document}